# DNA Sensing with Whispering Gallery Mode Microlasers

*Soraya Caixeiro[1,2]\*‡, Robert Dörrenhaus[3]‡, Anna Popczyk[1], Marcel Schubert[1], Stephanie Kath-Schorr[3]\*, Malte C. Gather[1,4]\**

[1]Department of Chemistry and Biochemistry, Humboldt Centre for Nano- and Biophotonics, Institute for Light and Matter, Greinstr. 4-6, 50939 Cologne, Germany

[2]Centre for Photonics and Photonic Materials, Department of Physics, University of Bath, Bath BA2 7AY, United Kingdom

[3]Department of Chemistry and Biochemistry, Institute of Organic Chemistry, Greinstr. 4, 50939 Cologne, Germany

[4]Centre of Biophotonics, SUPA School of Physics and Astronomy, University of St Andrews, St Andrews KY16 9SS, United Kingdom

‡ Authors contributed equally

\*Corresponding authors:
scc201@bath.ac.uk, skathsch@uni-koeln.de, malte.gather@uni-koeln.de

ABSTRACT. Nucleic acid sensing is crucial for advancing diagnostics, therapeutic monitoring and molecular biology research, by enabling the precise identification of DNA and RNA interactions. Here, we present an innovative sensing platform based on DNA-functionalized whispering gallery mode (WGM) microlasers. By correlating spectral shifts in laser emission to changes in refractive index, we demonstrate real-time detection of DNA hybridization and structural changes. The addition of gold nanoparticles to the DNA strands significantly enhances sensitivity, and labeling exclusively the sensing strand or a hairpin strand eliminates the need for secondary labeling of the target strand. We further show that ionic strength influences DNA compactness, and we introduce a hairpin-based system as a dual-purpose sensor and controlled release mechanism for potential drug delivery. This versatile WGM-based platform offers promise for sequence-specific nucleic acid sensing, multiplexed detection, and in vivo applications in diagnostics and cellular research.

**Introduction**

Nucleic acid chemistry is a fast-growing field with major implications for diagnostic and therapeutic applications as well as materials science[1–3]. A comprehensive understanding of the structure and dynamics of DNA and RNA is essential for a better understanding of their functions and for a deeper insight into their mechanisms of action[4]. In addition to the formation of double-stranded DNA helices, the diverse folding of RNA or single-stranded DNA (ssDNA) and the dynamic changes in their structures under different conditions are essential for the function of nucleic acids[5–7]. In particular oligonucleotide-based sensors are an expanding field of research and have proven to be a groundbreaking tool for prognosis and diagnosis[8–14].

Lasers are well suited for enhancing the sensitivity of a wide range of measurements, primarily due to their narrow emission spectra resulting from stimulated emission. These properties are sometimes overlooked, and lasers are often used simply as a bright, directed, and narrow band excitation source for various applications including excitation of fluorescence, Raman, surface plasmon resonance and others. Instead, lasers can also be employed as sensors in their own right[15–17], e.g. when their spectrum is used as a highly accurate external spectral ruler that shifts[18,19] and changes in intensity[20–22] through interaction of the laser with an analyte. This variant of laser-based sensing benefits considerably from the ability to make microscopic (volume $\ll (50~\mu m)^3$), low-cost, robust and biocompatible lasers. In this context, the use of whispering gallery mode (WGM) lasers is particularly attractive. WGM lasers often operate by optical pumping of dye molecules embedded in an otherwise transparent microsphere, made for instance of polystyrene, and rely on the optical confinement of the resulting emission inside the microsphere due to a refractive index (RI) contrast with the environment. Over the last decade, there has been a quickly growing body of work where such lasers have been directly integrated with biological matter, e.g. for tracking of cell migration[23–25] and sensing cellular forces[26,27], the latter even including non-contact measurements of local contractility in the living heart[19]. Looking at laser-based DNA sensing, double stranded DNA (dsDNA) intercalated with a fluorescent dye has been employed both as a laser gain medium[28] and as a sensing conduit, with sensing activated through staining after the DNA is hybridized, a process that occurs only when the base pairs between the strands match[29].

Gold nanoparticles (Au NPs) have a wide range of applications in optics, catalysis, biomedicine and sensing, owing to their controllable physiochemical properties, high chemical stability, good biocompatibility and excellent accessibility by wide-ranging surface functionalisation[30–32]. These unique characteristics make Au NPs valuable for sensing applications involving variations in interactions between nanoparticles with differences in parameters such as particle type, shape, relative position and number of particles[33,34] or even by assembling Au NPs with DNA to build complex structures with target-tailored functionalities[35,36]. Polystyrene microspheres coated with Au-NPs have been used to sense viral DNA using a fluorescently labelled single-stranded (ssDNA) capture oligonucleotide, this method relies on a fluorescently labelled ssDNA capture oligonucleotide, where fluorescence quenching occurs upon hybridization with the target DNA[8].

Beyond fluorescence-based sensing, Au NPs have also been used to locally enhance the electric field, therby improving WGM based detection of DNA[37-39]. Their ability to concentrate electromagnetic fields at the nanoscale enhances optical signals, increasing the sensitivity of biosensors. Over the past few decades, techniques for detecting short oligonucleotide sequences through in situ hybridization have expanded considerably[8,40-42].

Here, in order to gain insight into the structure and dynamics of oligonucleotides, we developed a novel method for sensing structural changes in DNA immobilized on a WGM microlaser that uses the minute local variation in RI cause by these structural changes. This approach exploits the unique optical properties of WGM microlasers and their ability to measure external RI. We further enhance the sensitivity of our measurement by the addition of Au NPs to the DNA, which allows for the specific sensing of short DNA fragments.

In the example used in our present study, the probe is based on a green-emitting, polystyrene-based WGM microlaser that is functionalized with ssDNA. Our model system consists of complementary-strand DNA (csDNA) strands and DNA hairpin forming strands. The polystyrene microlaser is covalently bound via a short linker to the strained cyclooctyne SCO-PEG3 modified 3'- end of a ssDNA sequence, which is modified with an amino group on its 5'-end and bound to a 2.2 nm diameter Au NP. Hybridization with csDNA leads to an increase of the RI in the immediate vicinity of the microlaser. This increase can be detected reliably through ensemble experiments on multiple lasers or via live analysis of an individual laser; the latter also provides insights into the hybridization dynamics. Moreover, we have developed a test system for a cleavable hairpin-based carrier that releases its Au NP only after the detection of specific DNA fragments.

In the future, with further investigation, our method could be adapted to detect the dynamics of ssDNA folding or can indicate hybridization or denaturation under different conditions. Our platform is highly versatile and can be adapted to detect different sequences of interest. Taken together, these characteristics make DNA functionalized WGM lasers a valuable tool for various applications in DNA sensing and analysis.

**Microlaser Surface modification with ssDNA and Au NPs**

The WGM microlasers used in this study are commercially available, monodisperse fluorescent polystyrene (PS) microspheres, approximately 11 µm in diameter, featuring carboxyl groups on their surface for functionalization with ssDNA, and containing a green-emitting fluorescent dye dispersed within the PS matrix. The DNA sequences used are provided in the Supporting Information (SI), Table S1. In brief, these sequences consist of a 22-base sequence modified at the 5'-end with an amine group for coupling to the carboxyl groups on the WGM microlasers or on Au NPs. Modifications at the 3'-end varied; some sequences were left unmodified, while others were conjugated with ~~strained cyclooctyne~~ SCO-PEG3 for Au NP attachment or tagged with Cy5 dye for conjugation confirmation, see Figure S1.

Carboxyl groups on both the microlasers and the Au NPs were activated using carbodiimide chemistry, facilitating the formation of amide bonds with the amine-modified DNA. To prevent non-specific binding, ethanolamine was used to block unreacted carboxyl groups. The DNA was then conjugated either to the Au NPs or directly to the microlasers.

To conjugate Au NPs, the Au NPs were covalently bound to the amino modification on the DNAs 5'-end and the microlasers were first modified with 11-azido-3,6,9-trioxaundecan-1-amine, allowing strain-promoted alkyne azide cycloaddition reaction (SPAAC) with the SCO-PEG3-terminated modified DNA, as described in the Methods, Section 2, in the SI and illustrated in Figure 1a.

To confirm successful modification of the microlaser surfaces with ssDNA, ssDNA tagged with the fluorescent dye Cy5 at their 3'-end was conjugated to the microlaser surface. Cy5 was selected for its red emission, which is well-separated from the absorption and emission spectra of the microlasers. Figure 1b, c and d show the green fluorescence from the microlasers, the red fluorescence from ssDNA conjugated to its surface, and an overlay of the two images, respectively. The clearly visible fluorescent ring at the microlaser surface confirms the successful conjugation of ssDNA.

To confirm the modification of ssDNA strands with Au NPs and their subsequent attachment to the microlasers, we used 40 nm diameter Au NPs. This size was chosen for its ease of visualization using scanning electron microscopy (SEM), as detailed in Section 2.6 of the Methods section in the SI. (Due to their small size and resolution limitations, 2.2 nm Au NPs were not visible under SEM, as shown in Figure S2.) All subsequent work used Au NPs with a diameter of 2.2 nm to avoid influencing DNA folding by NPs that are large compared to the DNA duplexes used, which have a calculated length of 6.6 nm for B-DNA duplexes. Figure 1d shows an SEM image of the smooth surface of an unmodified, carboxylated microlaser, while Figure 1e depicts a microlaser surface decorated with ssDNA conjugated with 40 nm Au NPs; a more zoomed in image is provided in Figure S3 of the SI.

As further evidence for successful NP functionalization, we look at the addition of ssDNA-conjugated with 40 nm Au NPs to a solution of azide-modified microlasers. Initially, the solution appeared red (Figure 1f, left tube) due to the plasmon resonance of the Au NPs dispersed in the solution. After 3 hours, most of the NP-conjugated ssDNA has reacted with the microlasers, which precipitated to the bottom of the container due to their size and weight. As a result, the solution became nearly transparent (Figure 1f, right tube).

Lastly, csDNA was introduced, matching the sequence of the ssDNA conjugated to the microlaser surface. To confirm conjugation, microlasers previously modified with the ssDNA were reacted with csDNA that was conjugated with Cy5 on its 3'-end. While microlasers conjugated with ssDNA alone showed minimal red fluorescence (Figure 1h), there was a distinct ring-like emission on the surface after reaction with csDNA (Figure 1i), confirming successful hybridization to dsDNA.

**Lasing and Sensing DNA with microlasers**

The fluorescent molecules embedded within the PS microspheres can provide optical gain. When pumped with pulsed laser light above a threshold energy density of approximately 125 µJ/cm², as established by previous studies[18], the microspheres therefore emit laser light, and their emission spectra are dominated by a series of WGM lasing peaks. The emission spectrum of the lasers is characterized by a series of sharp peaks associated with alternating transverse-electric (TE) and transverse-magnetic (TM) modes, where the electric field of the light in the WGM is oriented either parallel (TE) or perpendicular (TM) to the surface. The exact wavelength of these peaks strongly depends on both the microlaser size and the RI in the immediate vicinity of the laser. A typical experimentally observed lasing spectrum is shown in Figure 2a. Figures 2b and 2c show simulated emission spectra for a microlaser embedded in two media of different RI (details on how these spectra were generated can be found in Section 2.7 of the Methods section in the SI). For the medium with higher RI, a red shift in the position of all peaks/modes is observed. By carefully analysing the modal positions and knowing the internal RI of the microlaser, it is possible to determine both the microlaser size and the external RI, following a routine described e.g. in Ref. 43. Using this approach, the external RI for the microlaser in Figure 2a is determined to be 1.339.

Figures 2d-h summarizes the external RI for N>40 microlasers for each successive step of surface functionalization, measured in the same buffered solution, clearly demonstrating a trend of increasing RI. As expected, the carboxylated microlasers exhibit the lowest external RI (Figure 2d), serving as the baseline for later modifications. Upon addition of ssDNA, we observe a modest increase in mean external RI of just 0.0013 RIU, slightly smaller than the standard deviation of the mean, which is around 0.002 RIU and is not statistically significant (Figure 2e). Similarly, the transition from ssDNA to helical B-form dsDNA through hybridization does not produce a statistically significant change in RI (Figure 2f), suggesting that the structural alterations associated with hybridization are not captured within the resolution limit of the ensemble measurement.

A more pronounced effect is observed when ssDNA is conjugated with Au NPs at the 5'-end; in this scenario, the RI increases by 0.004 RIU (Figure 2g) compared to ssDNA without Au NPs and by 0.005 RIU compared to the bare carboxylated microlasers, resulting in statistically significant differences in refractive index distributions (p <0.05). The larger standard deviation of 0.004 RIU likely reflects increased sample heterogeneity.

Compared to our initial experiment with unmodified DNA, the RI change from ssDNA-Au NPs to dsDNA-Au NPs conjugates is relatively small but statistically significant (p <0.05), with an average increase of 0.002 RIU (Figure 2h). Overall, these results demonstrate that the attachment of gold nanoparticles is important for producing measurable and significant changes in external RI upon DNA binding to microlasers.

WGM modes decay exponentially away from the microlasers or resonators surface[44], for the resonator geometry described here the 1/e extension for a typical TE mode is approximately 120 nm[18]. The observed increase in average external refractive index is influenced by the full extension, with the highest sensitivity occurring near the surface where the field overlap is strongest.

Thus, the refractive index measurement does not represent the intrinsic refractive index of DNA or Au NPs alone but rather their combined effect with the buffer medium. When the DNA helix forms, bringing Au NPs closer to the surface, the increased overlap with the evanescent field leads to a higher measured refractive index.

Next, we examined how ionic strength and buffer concentration affect the RI measured by microlasers. Microlasers modified with ssDNA-Au NPs were hybridized with complementary csDNA in buffers of varying concentrations. Since buffer concentration correlates with ionic strength, this approach enabled us to systematically investigate the role of ion shielding on DNA hybridization and compaction. The increased ionic strength corresponds to higher cation concentrations, which effectively shield the negatively charged phosphate groups along the DNA backbone and thus lead to a more stable structure[45–47]. Salt ions weaken the electrostatic repulsion of the phosphates and allow the DNA strands to approach one another, promoting a more compact structure[48,49].

Figure 3 presents the RI distributions for N>40 microlasers under different buffer conditions. Figure 3a shows ssDNA-Au-NP functionalized microlasers and provides a baseline for sample-to-sample variations. Upon hybridization with csDNA, a notable upshift in the RI distribution is observed, as shown in Figure 3b for a buffer concentration of 0.01 M. We attribute the increase in RI to the hybridization bringing Au NPs closer to the microlaser surface due to formation of the compact B-form double helix of the DNA[50].

At a higher buffer concentration of 0.1 M, we observed a further increase in RI, with an average change of 0.004 RIU relative to the 0.01 M buffer concentration ($p < 0.05$). This finding further confirms that an increase in ionic strength increases the compactness of the dsDNA helix, facilitates closer proximity of the Au NPs to the microlaser surface, and thereby amplifies the RI perceived by the lasing modes.

**Real-time DNA hybridization sensing with microlasers**

Having demonstrated that microlasers functionalized with Au-NPs provide an effective platform for sensing DNA hybridization and distinguishing different ionic strengths in ensembles of microlasers, we now explore detection of DNA hybridization of strands with varying lengths. To do so, we follow the RI change in the vicinity of individual microlasers due to hybridization events in real-time. Specifically, a microlaser decorated with ssDNA conjugated to a NP at its 5'-terminus is continuously monitored for spectral changes, while csDNA is added in solution (Figure 4a). In Figure 4b, we show a TE/TM pair of WGM modes, and follow their shift over the course of the experiment. The corresponding changes in RI, calculated from the spectral shifts, are plotted in

Figure 4c. (Additional lasing modes are present in the spectrum but are not displayed for simplicity.)

Before the addition of csDNA, the spectra and the calculated corresponding RIs remain unchanged, as indicated by the blue and lilac points and spectra in Figures 4b and 4c, respectively. Upon introduction of csDNA at time zero, a red shift in both the TM and TE modes is observed, corresponding to an increase in the RI. Once the system reaches equilibrium, the spectra stabilize, as represented by the green and pink points and spectra in Figures 4b and 4c, respectively. As these experiments are not influenced by inherent variability between microlasers (e.g., in size, internal RI, coverage, etc.), they can monitor the binding of molecules in real time and provide drastically improved sensitivity. As such it is also possible to measure the hybridization of DNA in the absence of Au-NPs (SI Figure S4). To further validate our findings, we also perform measurements on a control sample consisting of carboxylated microlasers without ssDNA attached to the surface. Upon addition of csDNA, no significant increase in refractive index is observed (SI Figure S5), confirming that the observed RI shift in our primary experiments is specifically due to DNA hybridization rather than nonspecific interactions.

We also explored different hybridization configurations, where the Au-NPs were on the ssDNA, on csDNA, on both or neither, real-time hybridization curves were obtained and the change in refractive index are compared in SI Figure S6. Notably, the refractive index change was found to be most pronounced when the Au-NPs were attached to the ssDNA, which is the primary focus of this study.

**DNA hairpin fold resolving sensing**

Finally, we developed an inverted microlaser-based test system for DNA sensing, which may also be employed in the future as target-sensitive hairpin-based drug delivery system[51]. To explore this, we employed a six nucleotide hairpin (HP) sequence construct forming a 4 nucleotide loop[51] and modified with Au-NPs by amide coupling of COOH-modified Au-NPs with amino-linker functionalized hairpin-DNA strands on the 5'-end, as detailed in the Method section of the SI. The baseline RI distribution for microlasers conjugated with a ssDNA (Figure 5a) agrees with the RI values previously seen in samples with ssDNA-NPs (Figure 2g) and in hybridized systems containing csDNA (Figure 2h). When adding csDNA, we expected that csDNA replaces the HP, resulting in a decrease in RI by removal of the Au-NPs. However, contrary to this expectation, an increase in RI was observed. Binding energy predictions (see SI, Figure S7) and melting temperature measurements (See SI, Figure S8 and S9) indicate that at room temperature (RT), the high binding energies—derived from two sets of three matching base pairs—maintain the stability of the test system. This causes a contraction in the overall construct length, bringing Au-NPs closer to the microlaser surface, as observed when the csDNA binds, and consequently increasing the measured RI.

After performing hybridization at elevated temperatures of 37 °C and 60 °C for one hour, the anticipated substitution of HP-Au-NPs was observed, i.e. microlaser measurements at room temperature (RT) showed a decrease in local RI decreased. At 37 °C (Figure 5d), partial removal of HP-Au-NPs was observed, with the RI remaining significantly above that of ssDNA-conjugated microlasers. At 60 °C (Figure 5e), nearly complete removal was achieved, and the RI returned to the range observed for unmodified samples (compare Figure 5a).

**Conclusion**

In summary, we demonstrated an innovative approach to nucleic acid sensing using WGM microlasers functionalized with DNA and Au NPs. Our method leverages real-time monitoring of RI changes near the microlaser surface, providing a sensitive platform for detecting DNA hybridization and structural dynamics. It enables detection of sequences in two distinct configurations: either by immobilizing the target complementary strand on the laser surface with an Au NP attached to its 5'-end, or by binding a bare strand to the laser surface and complementing this with an Au NP-modified hairpin strand. In both scenarios, we successfully detect an unmodified target DNA sequence. Compared to many fluorescence-based methods, our technique requires only minor 5'-end modifications on the DNA, effectively circumventing the issue of fluorophore quenching[52]. The addition of Au NPs significantly enhances the sensitivity of this WGM-based detection, as RI changes become more pronounced with nanoparticle proximity. Additionally, we explored the influence of ionic strength on hybridization, showing that variations in ion concentration affect DNA compactness and, consequently, RI shifts. Furthermore, a hairpin-based test system has been developed to assess DNA hybridization and as a potential controlled release mechanism, enabling potential applications in targeted drug delivery.

Our findings pave the way for broader applications of WGM microlaser sensors, particularly in areas requiring precise detection of nucleic acid interactions, such as diagnostics and environmental monitoring. This platform enables the use of customisable DNA sequences tailored to bind various targets, e.g. nucleic acid sequences by base pairing, to small molecules or peptide/proteins using aptamer sequences.

Finally, our platform provides a highly adaptable basis for sequence-specific detection systems. Understanding conformational changes in nucleic acids and investigating biomechanical forces within cells is increasingly important for elucidating cellular functions and molecular interactions.

The structural transitions in DNA and RNA—such as folding, hybridization, and denaturation—are critical for processes such as transcription, replication, and signaling. Integration of this system with different capture strands has the potential to enhance multiplexed sensing capabilities, enabling simultaneous monitoring of multiple analytes in complex biological samples, potentially redefining current approaches to nucleic acid detection and analysis.

Figure 1. Preparation and optical characterization of the surface coating with gold nanoparticle (Au-NP) functionalized DNA on microlasers. (A) Schematic of the nanoparticle ssDNA conjuga-

tion to a carboxyl functionalized microlaser. (B) Electron microscopy image of a carboxyl functionalized microlaser. Scale bar is 2 µm. (C) Electron microscopy image of microlaser decorated with DNA functionalized with 40 nm diameter Au-NPs. Scale bars are 2 µm and 500 nm (inset). (E) Green fluorescence from microlaser functionalized with ssDNA with Cy5 dye. (F) Red fluorescence of same microlaser. (G) Overlay of green and red emission. (H-J) Microlaser with dye-free ssDNA on surface shows no red fluorescence (H). The hybridization with csDNA containing Cy5 illustrated in (I) leads to the appearance of a red ring in the fluorescence image (J). Scale bars for E, F, G, H and J are 20 µm.

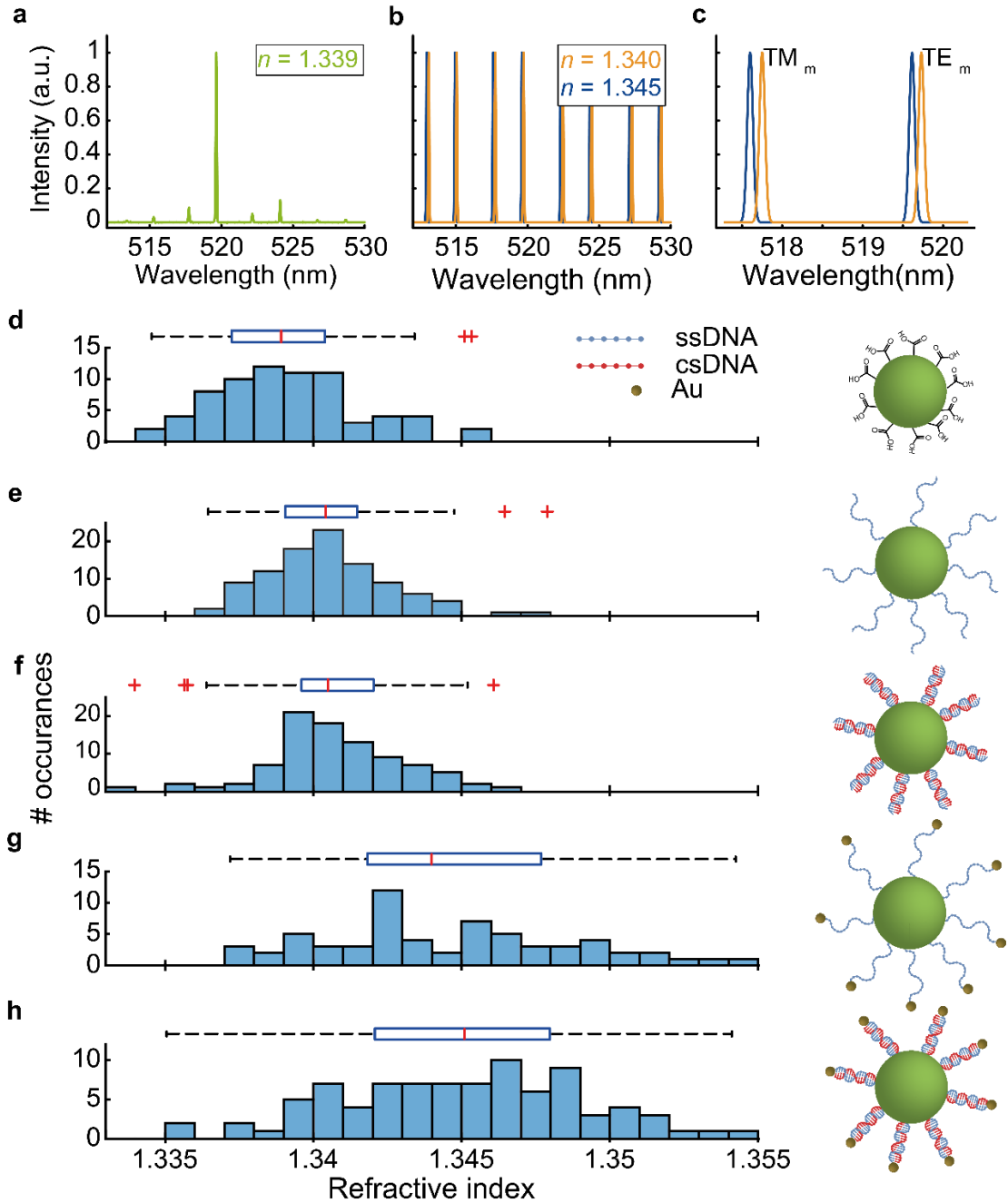

Figure 2. Refractive index sensing of DNA surface modifications. (A) Representative lasing spectra from a microlaser in buffered solution, along with the fitted refractive index. (B) Simulated lasing spectra for microlasers embedded in media of two different refractive indices, $n$ =

1.34 (orange) and $n = 1.345$ (blue). (C) Zoom in of the simulated spectra, highlighting the spectral shifts across different refractive indices and polarization, for azimuthal mode number $m = 105$. Histograms of external refractive indices calculated from measured laser spectra, with a corresponding box plot, showing (D) carboxyl-functionalized microlasers (N=71), (E) microlasers conjugated with ssDNA (N=99), (F) microlasers conjugated with dsDNA (N=90), (G) microlasers conjugated with ssDNA and Au NPs (N=65), and (H) microlasers conjugated with dsDNA and Au NPs (N=91). The illustrations to the right of each histogram depict the configuration for each case.

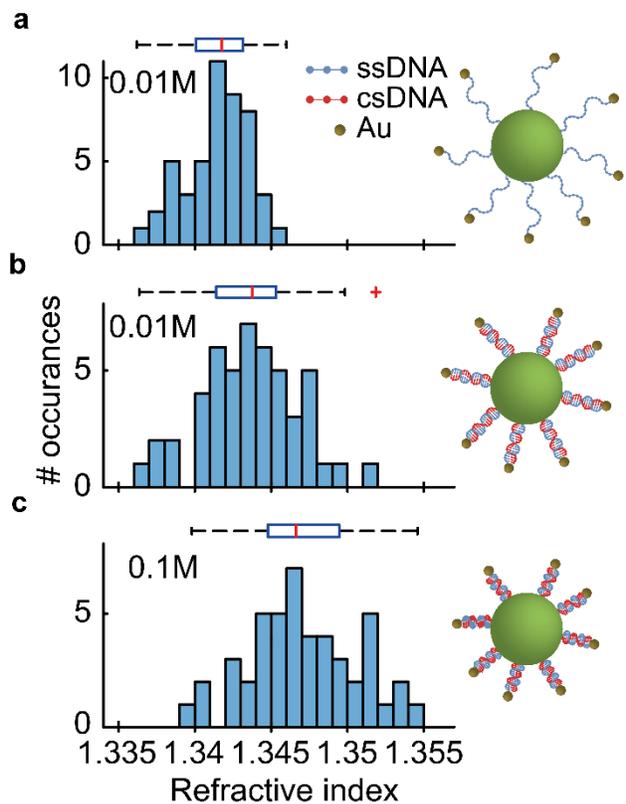

Figure 3. Detection of DNA compression at varying ionic strengths of buffer solutions. Histograms of refractive index of (A) ssDNA-Au-NP functionalized microlasers in 0.01 M concentration buffer solution (N=48), and microlasers following hybridization with csDNA in (B) 0.01 M (N=47) and (C) 0.1 M (N=49) concentration buffer solution.

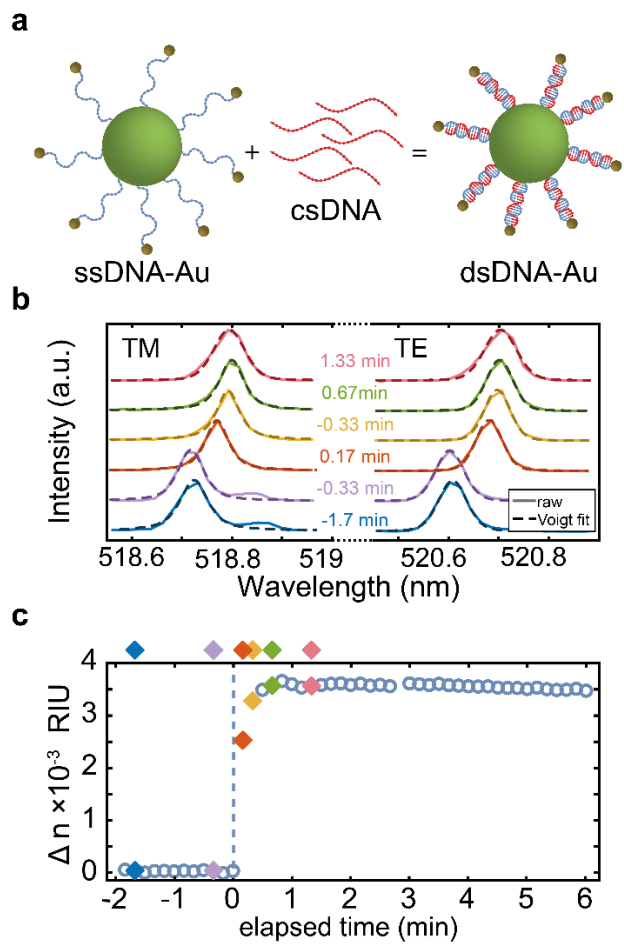

Figure 4. Real-time detection of DNA hybridization on the surface of a microlaser. (A) Schematic representation of DNA hybridization on single-stranded DNA (ssDNA) functionalized with Au nanoparticles (Au NPs). (B) Spectral shifts of a selected TM and TE mode from a single microlaser at different time points during the reaction. (C) Transient refractive index change calculated from the microlaser spectra acquired during DNA hybridization on the laser surface. Time zero (t = 0) indicates the moment csDNA was introduced into the solution. Filled diamonds indicate specific time points with corresponding spectra shown in panel (B).

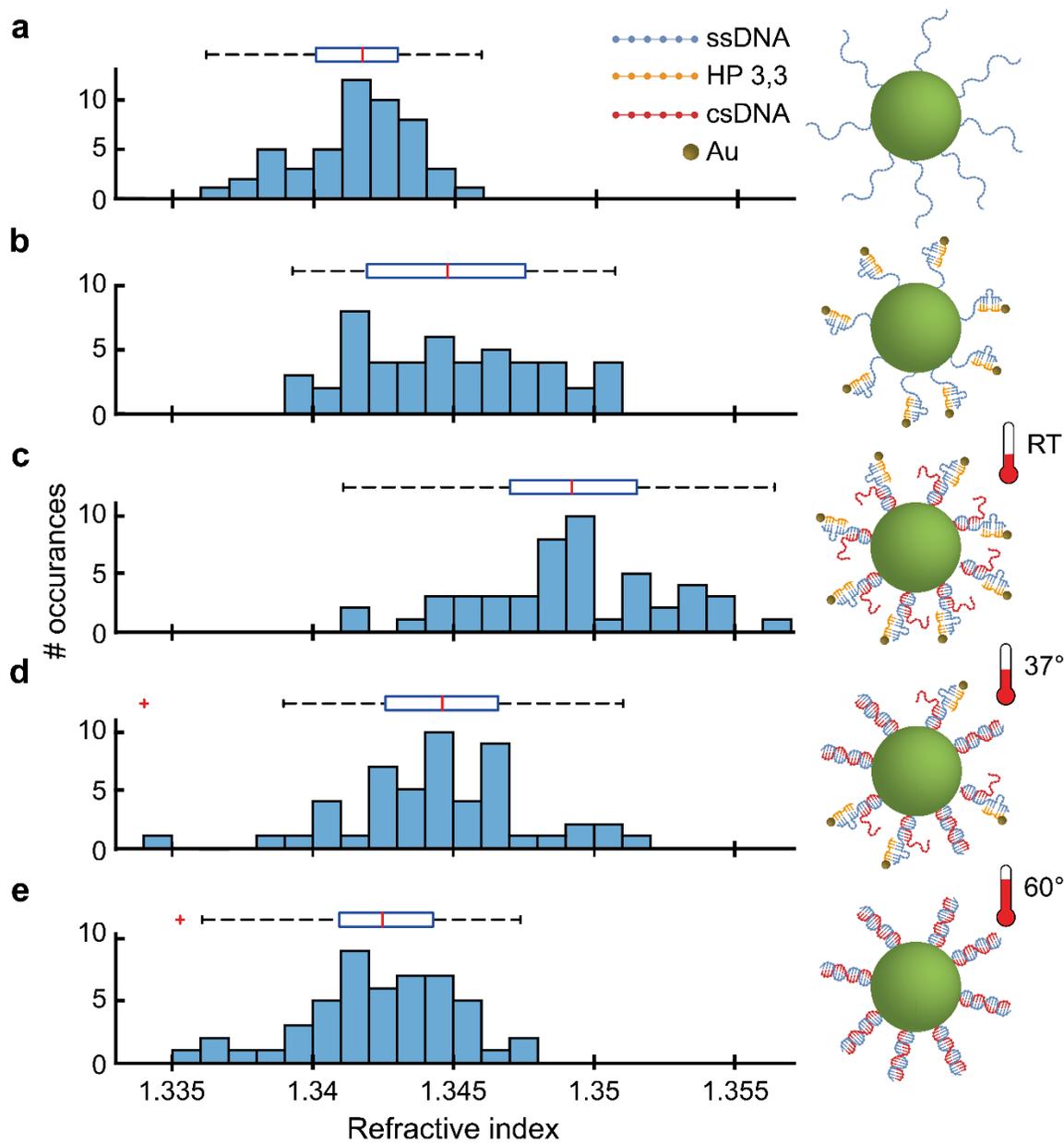

Figure 5: Refractive index change upon addition and substitution of hairpin DNA. (A) Refractive index histogram for microlasers functionalized with ssDNA. (B) Refractive index histogram after addition of the HP33 hairpin. (C-E) Refractive index histograms after substitution was performed by addition of csDNA for 1 h reaction at (C) room temperature (RT), (D) 37 °C, and (E) 60 °C. Refractive index measurements were performed at RT in all cases.

For Table of Contents Only:

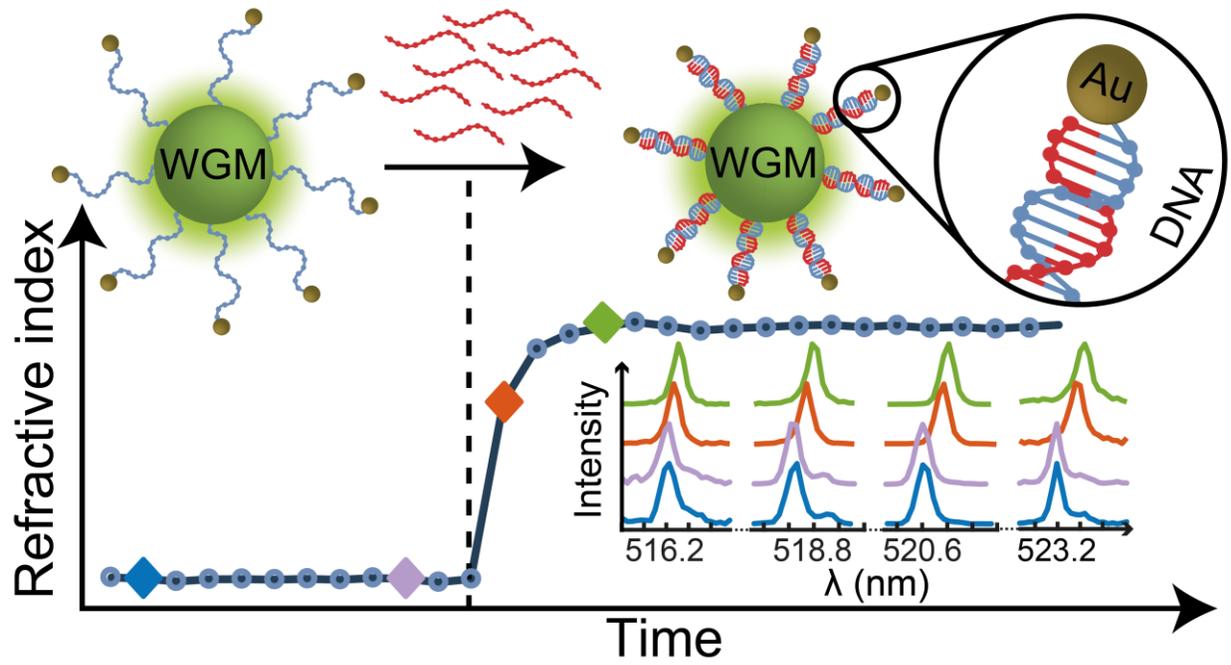

## ASSOCIATED CONTENT

**Supporting Information**. Detailed experimental information can be found in the SI. The following files are available free of charge.

Supporting Information (PDF)

## Author Contributions

The manuscript was written through contributions of all authors. All authors have given approval to the final version of the manuscript. ‡These authors contributed equally. (match statement to author names with a symbol)

## Funding Sources

This work received financial support from the Humboldt Foundation (Alexander von Humboldt professorship to M.C.G.).

## ACKNOWLEDGMENT

The authors wish to thank Prof. Jan Riemer for fruitful discussions.

# DNA Sensing with Whispering Gallery Mode Microlasers

# Supplementary Information

*Soraya Caixeiro[1,2]*‡, Robert Dörrenhaus[3]‡, Anna Popczyk[1], Marcel Schubert[1], Stephanie Kath-Schorr[3]*, Malte C. Gather[1,4]**

[1]Department of Chemistry and Biochemistry, Humboldt Centre for Nano- and Biophotonics, Institute for Light and Matter, Greinstr. 4-6, 50939 Cologne, Germany

[2]Centre for Photonics and Photonic Materials, Department of Physics, University of Bath, Bath BA2 7AY, United Kingdom

[3]Department of Chemistry and Biochemistry, Institute of Organic Chemistry, Greinstr. 4, 50939 Cologne, Germany

[4]Centre of Biophotonics, SUPA School of Physics and Astronomy, University of St Andrews, St Andrews KY16 9SS, United Kingdom

‡ Authors contributed equally

*Corresponding authors:
scc201@bath.ac.uk, skathsch@uni-koeln.de, malte.gather@uni-koeln.de

## 1. Materials:

All chemicals were purchased from *ABCR, BLD Pharm, Fisher Scientific, Roth, Sigma Aldrich, TCI*. DNA strands were purchased from *Biomers*. A list of used DNA strands is shown in the table below. Carboxylated spherical gold nanoparticles (product Number: C11-2-2-TC-DIH-50-1 for 2.2 nm and C11-40-MC-DIH-50-1 for 40 nm) were purchased from *Nanopartz* and polystyrene beads from *Polysciences Inc*. (Item Code: 18142-2, Fluoresbrite® YG Carboxylate Microspheres 10.00 µm).

Table S1: Used DNA strands and modifications.

| DNA | 5'-end | Sequence | 3'end |
|---|---|---|---|
| ssDNA-Cy5 | Amino C6 | TCA ACA TCA GTC TGA TAA GCT A | Cy5 |
| csDNA | - | TAG CTT ATC AGA CTG ATG TTG A | - |
| ssDNA | Amino C6 | TCA ACA TCA GTC TGA TAA GCT A | - |
| csDNA-Cy5 | Cy5 | TAG CTT ATC AGA CTG ATG TTG A | - |
| ssDNA-SCO | Amino C6 | TCA ACA TCA GTC TGA TAA GCT A | SCO-PEG3 |
| csDNA-Au | Amino C6 | TAG CTT ATC AGA CTG ATG TTG A | - |
| HP-33 | Amino C6 | AGC CAG | - |
| RandomDNA | - | GTT AAC GAG TTC AAC TCC AGA C | - |

**Reagents:**

Stock solutions were prepared using distilled or deionized water unless otherwise indicated. To prepare the 0.1 M carbonate buffer, 0.1 M $NaCO_3$ was added to 0.1 M $NaHCO_3$ until pH 9.6 was reached.

For 0.1 M MES Buffer, 19.2 g of MES free acid (MW 195.2 g/mol) was dissolved in 900 ml of pure water, then titrated to the desired pH (5.2-6.0) with 1N NaOH, and the volume was filled up to 1000 ml with pure water.

The 2% carbodiimide solution consisted of 2% 1-(3-dimethylaminopropyl)-3-ethyl carbodiimide hydrochloride dissolved in 0.1 M MES buffer. This was always freshly prepared and used within 15 minutes of preparation.

To prepare 0.2 M borate buffer, 1 M NaOH was added to boric acid until pH 8.5 was reached.

The 0.25 M ethanolamine solution was prepared by adding 20 µl of ethanolamine (2-aminoethanol) to 1.3 ml of borate buffer.

For the storage buffer, a 0.01 M phosphate buffer (pH 7.4) was prepared with 0.1% sodium azide and 5% glycerol. This involved preparing a 0.1 M stock of sodium phosphate monobasic ($NaH_2PO_3$), 13.8 g/l, monohydrate (MW 138.0 g/mol), and a 0.1 M stock of sodium phosphate dibasic ($Na_2HPO_3$), 26.8 g/l, heptahydrate (MW 268.0 g/mol). 19.0 ml of $NaH_2PO_3$ solution was mixed with 81.0 ml of $Na_2HPO_3$ solution to yield pH 7.4. After adding 50 ml of glycerol and 1.0 g of sodium azide, the mixture was diluted to a final volume of 1.0 l.

For melting curve measurements, 100 ml of 0.01 M phosphate-buffered saline (PBS) with 0.01 M NaCl was prepared. 8.2 ml of a 0.1 M sodium phosphate dibasic ($Na_2HPO_3$) stock solution and 1.8 ml of 0.1 M sodium phosphate monobasic ($NaH_2PO_3$) stock solution were combined in a flask with approximately 80 ml of distilled water. pH was adjusted to 7.0 using HCl or NaOH as needed. Then, 0.0584 g of NaCl was added to the mixture and the total volume was brought to 100 ml with distilled water. The solution was thoroughly mixed until all components were dissolved.

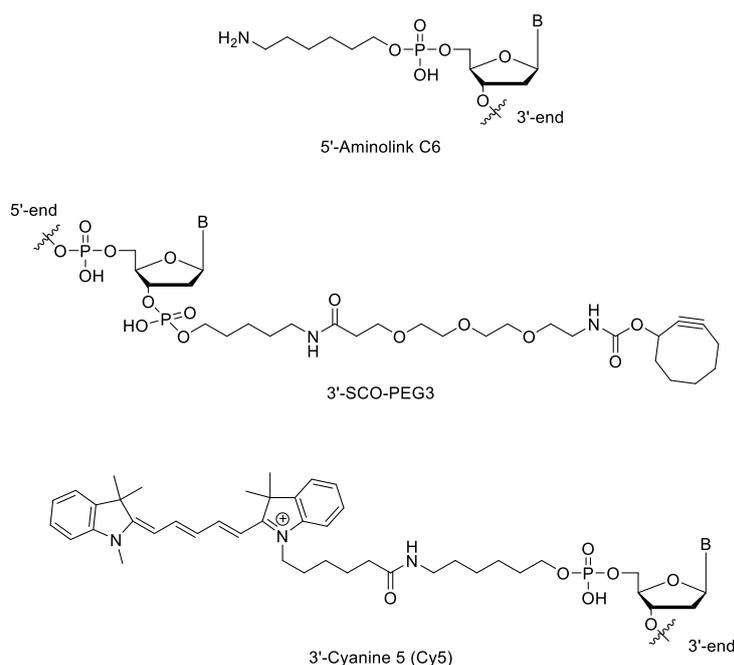

Figure S1: 5'-Aminolink (top), 3'-SCO-PEG3 Linker (middle) and 3'-Cy5 modification (bottom).

# 2. Methods
## 2.1. Microlaser functionalisation

General procedure:

10 µl of 2.5% carboxylated microparticles were placed into an Eppendorf centrifuge tube (1.5-1.9 ml capacity). 200 µl of 0.1 M carbonate buffer were added, then the mixture was centrifuged for 5-6 minutes in a microcentrifuge at 10,000 rpm. The supernatant was carefully removed and discarded using a Pasteur pipette. This process was repeated once. To resuspend the pellet, the tube was filled with half the mentioned amount and capped, vortexed, then filled to the written amount. The pellet was resuspended in 100 µl of 0.1 M MES buffer, centrifuged for 5-6 minutes, and the supernatant was carefully removed and discarded. This step was repeated twice more.
Next, the pellet was resuspended in 100 µl of 0.1 M MES buffer and 10 µl of 2% carbodiimide solution were added dropwise. The mixture was mixed for 10-15 minutes at room temperature, then centrifuged for 5-6 minutes and the supernatant was discarded. The pellet was resuspended in 100 µl of 0.1 M MES buffer, centrifuged, and the supernatant was discarded. This step was repeated twice more to remove unreacted carbodiimide.
The pellet was then resuspended in 100 µl of 0.2 M borate buffer. 7.5 µl of 100 pmol/µl DNA solution with one end amino-C6 functionalisation (or 2 µl of 11-azido-3,6,9-trioxaundecan-1-amine for azide functionalization) were added. The mixture was gently mixed overnight at room temperature on an end-to-end mixer. It was then centrifuged for 10 minutes, and the supernatant was removed. The volume of the supernatant was noted and stored for residue measurements by UV absorption. The amount of substance was calculated and subtracted from the starting amount.
The pellet was resuspended in 100 µl of 0.2 M borate buffer and 10 µl of 0.25 M ethanolamine were added. The mixture was gently mixed for 30 minutes with an end-to-end mixer to block unreacted sites of the nanoparticles. It was then centrifuged for 5-6 minutes and the supernatant was discarded. The pellet was resuspended in 100 µl of 0.2 M borate buffer, centrifuged, and the supernatant was discarded. This step was repeated twice more.
Finally, the pellet was resuspended in 100 µl of 0.01 M storage buffer. The bead concentration was now 0.25%.

## 2.2. Functionalization of Gold nanoparticles

### 2.2 nm Au-NPs:

10 µl of 2.5 mg/ml carboxylated colloid spherical gold nanoparticles (2.2 nm diameter) were placed into an Eppendorf centrifuge tube with a capacity of 1.5-1.9 ml. Then, 80 µl of 0.1 M MES buffer were added to the tube followed by the dropwise addition of 10 µl of 2% carbodiimide solution. The contents were mixed for 10-15 minutes at room temperature.
Subsequently, 10 µl of a 100 pmol/µl DNA solution, which contained DNA with a 3'-modification of SCO-PEG3 and a 5'-modification of Amino-C6, were added. The mixture was gently mixed overnight at room temperature on an end-to-end mixer. The solution was then used without further workup.

**40 nm Au-NPs:**

10 µl of 2.5 mg/ml carboxylated colloid spherical gold nanoparticles (40 nm diameter) were placed into an Eppendorf centrifuge tube with a capacity of 1.5-1.9 ml. 80 µl of 0.1 M MES buffer were added to the tube, followed by the dropwise addition of 10 µl of 2% carbodiimide solution. The mixture was then mixed for 10-15 minutes at room temperature.

After mixing, the solution was centrifuged for 10 minutes. The supernatant was carefully removed and discarded. The resulting pellet was resuspended in 100 µl of 0.1 M MES buffer. To this suspension, 10 µl of 100 pmol/µl DNA solution (containing DNA with 3'-modification SCO-PEG3 and 5'-modification Amino-C6) were added. The mixture was then gently mixed overnight at room temperature on an end-to-end mixer.

Following the overnight incubation, the solution was centrifuged for 10 minutes. The supernatant was removed and discarded, and the pellet was resuspended in 100 µl of 0.1 M MES buffer. This centrifugation, supernatant removal, and resuspension process was repeated once more to ensure thorough washing of the nanoparticles.

### 2.3. Strain-promoted Azide-Alkyne Click reaction (SPAAC) for functionalization of bead-DNA-Au

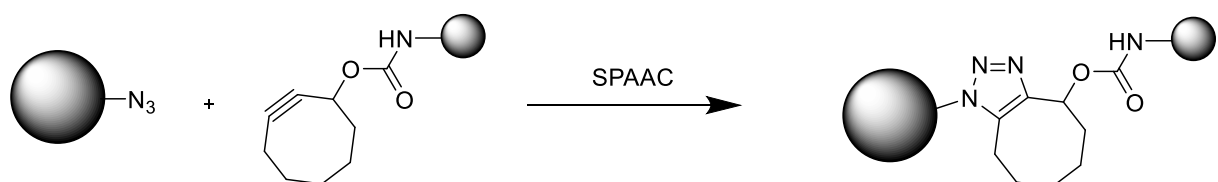

Azide functionalized beads suspended in storage buffer were prepared as described in the general procedure, resulting in a volume of 100 µl of 0.25% beads in suspension. Approximately 100 µl of freshly prepared Au-NP suspension, containing 0.025 mg of Au, was added to the bead suspension. The mixture was gently shaken overnight at 37°C using an Eppendorf Thermomixer comfort at 300 rpm.

Following incubation, the mixture was centrifuged for 10 minutes. The pellet was then resuspended in 100 µl of 0.1 M MES buffer, and centrifugation was repeated for 5-6 minutes, after which the supernatant was removed and discarded. The pellet was resuspended in 100 µl of 0.2 M borate buffer, and centrifugation was conducted again for 5-6 minutes, with the supernatant removed and discarded. This step was repeated two more times.

Finally, the pellet was resuspended in 100 µl of 0.01 M storage buffer.

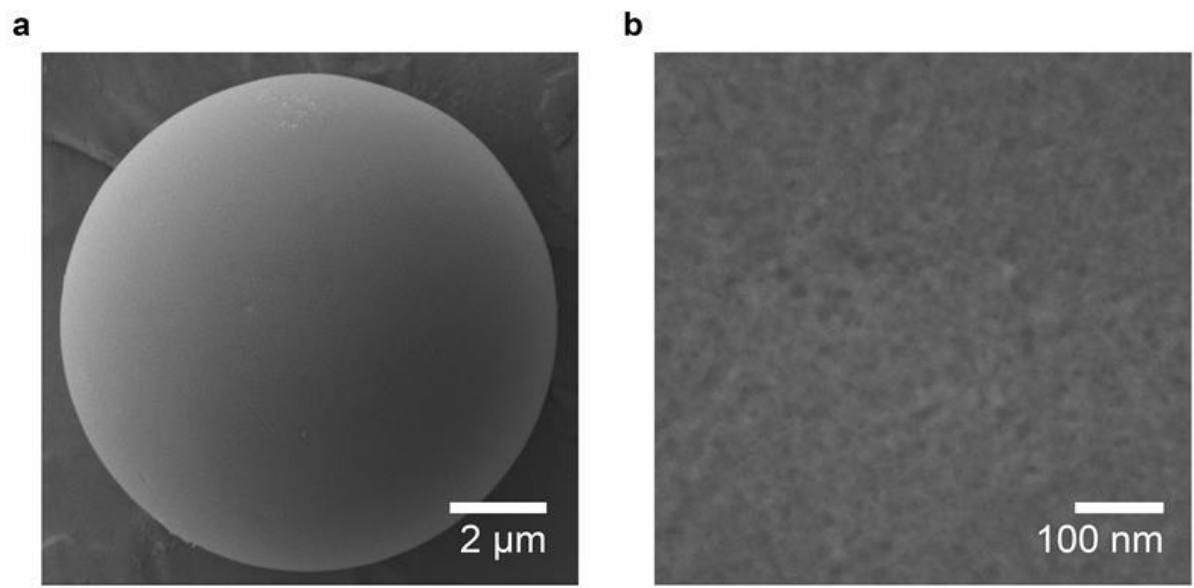

Figure S2 Electron microscopy image of microlaser decorated with DNA functionalized with 2 nm diameter Au-NPs.

Table S2: Measured and calculated analytical data provided by the suppliers of the functionalized spherical gold nanoparticles and the carboxylated microspheres used in this work.

| Functionalized Spherical Gold Nanoparticles | | | Carboxylate Microspheres |
|---|---|---|---|
| Product No. | C11-2.2-TC-DIH-50-1 | C11-40-MC-TC-DIH-50-1 | 18142-2 |
| Solution | DI water | DI water | DI water |
| Diameter (measured) | 40 nm | 2.2 nm | 10.6 µm |
| Size Dispersity %PDI | 20% | 7% | 4% |
| SPR Abs. (measured) | 56 OD | 78 OD | |
| SPR peak (measured) | 505 nm | 526 nm | |
| Concentration (calc.) | $2.61 \times 10^{16}$ nps/ml | $6.05 \times 10^{12}$ nps/ml | $4.55 \times 10^{7}$ ps/ml |

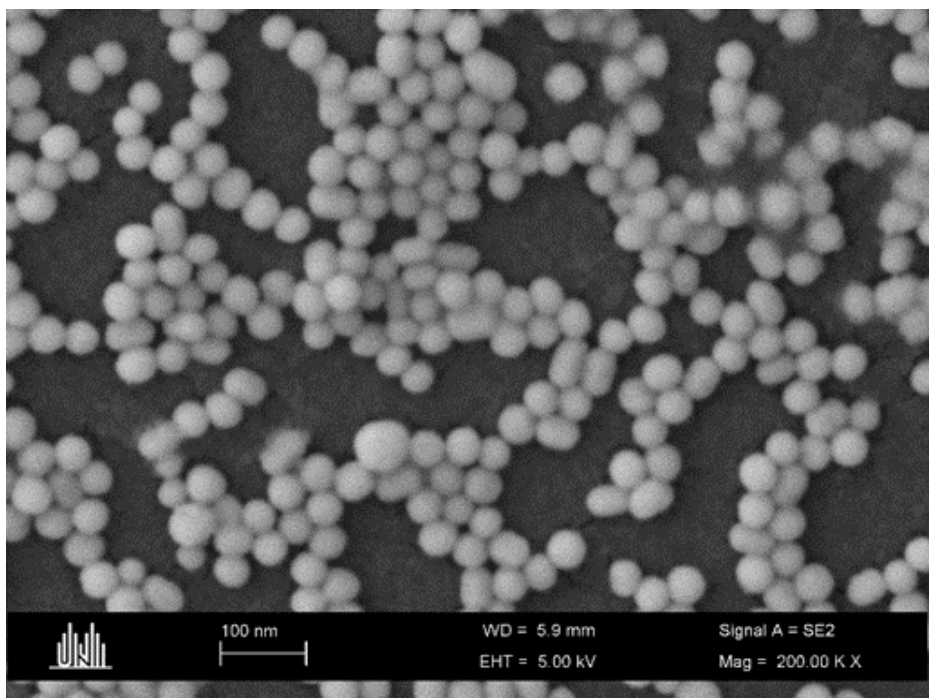

Figure S3 Electron microscopy image of 40 nm Au NPs.

## 2.4. Hybridization

The sample was vortexed for resuspension. Subsequently, 10 µl of complementary strand DNA (or the calculated amount determined via nanodrop in a previous step) were added to the sample. The mixture was mixed at 300 rpm in an Eppendorf Thermomixer comfort for 1 h to ensure proper interaction between the components.
After mixing, the solution was centrifuged for 5-6 minutes, and the supernatant was carefully removed and discarded. The resulting pellet was then resuspended in 100 µl of 0.01 M storage buffer. This centrifugation and supernatant removal process was repeated three more times to thoroughly wash the pellet. Finally, the pellet was resuspended in 100 µl of 0.01 M storage buffer.

## 2.5. Hairpin experiments

DNA modified beads were prepared as described in the general procedure. Hairpin strands were then modified with 2.2 nm Au-NPs following the specified procedure. For each planned experiment, 10 µl of a 0.25% bead suspension modified with ssDNA was placed in an Eppendorf centrifuge tube (1.5-1.9 ml capacity). Then, 90 µl of storage buffer were added, followed by 10 µl of a 10 µM suspension of HP-Au or csDNA-Au. The mixture was gently shaken in at 300 rpm for 60 minutes at 37°C. After incubation, the solution was centrifuged for 5-6 minutes, and the supernatant was removed and discarded. The resulting pellet was resuspended in 100 µl of 0.01 M storage buffer. These washing steps were repeated three more times to ensure thorough washing. The pellet was then resuspended in 100 µl of 0.01 M storage buffer (or 0.1 M storage buffer or other concentrations for concentration dependency studies). A sample of HP-Au hybridized to the bead with ssDNA was kept for reference.

Next, 10 µl of a 10 µM solution of csDNA were added to the suspension. The mixture was shaken in a Thermomix for 60 minutes at room temperature or at elevated temperatures. Following this incubation, the solution was centrifuged for 5-6 minutes, and the supernatant was removed and discarded. The pellet was resuspended in 100 µl of 0.01 M storage buffer. This centrifugation, supernatant removal, and resuspension process was repeated three more times.

Finally, the pellet was resuspended in 100 µl of 0.01 M storage buffer.

### 2.6. Imaging and laser spectroscopy of microlasers

The surface morphology of the microlasers was characterized using a Zeiss Neon40 Cross Beam Scanning Electron Microscope (SEM). Microlaser solutions were washed, diluted in deionized water, and carefully deposited onto an electron microscope stub layered with carbon tape. To mitigate charging effects during imaging, the samples were sputter-coated with a gold layer approximately 5–10 nm thick.

The microlasers were optically imaged using an inverted widefield optical microscope (Nikon Ti2) equipped with epifluorescence and differential interference contrast (DIC) capabilities. A 100x oil immersion objective (Nikon Plan Apo VC, NA 1.4) was used for imaging. They were functionalised as described in the methods above and imaged in cover-slip-bottom petri dish (ibidi) in storage buffer.

The microlasers were optically excited using a Q-switched, diode-pumped solid-state laser (Alphalas) operating at a wavelength of 473 nm, with a pulse width of 1.5 ns and a repetition rate set to 100 Hz. Depending on the functionalisation, the laser pulse energy ranged from 0.4 nJ to 6 nJ and was coupled into the objective via a dichroic mirror. This corresponded to a pump fluence of 0.2–2.9 mJ/cm², with an excitation spot with a diameter of 25 µm.

The emission from the microlasers was collected by the same objective and separated from the pump light via the dichroic mirror. The emission was relayed to a spectrometer (Andor Shamrock 500i and Andor Newton DU971P-BV) and a cooled sCMOS camera (Hamamatsu Orca Flash 4.0) through a series of lenses and dichroic beam splitters. The excitation and spectroscopy components were custom-built and integrated into a Nikon microscope platform.

Fluorescence imaging of the microlasers and DNA coverage was performed using the same setup, with an LED illumination system (CoolLED pE-4000) in the epifluorescence port. The microlaser fluorescence was imaged by setting the LED output to 460 nm and collecting fluorescence in the 520–535 nm wavelength range. For imaging the Cy5 dye, the LED was set to 635 nm, and fluorescence with wavelengths above 660 nm collected by the sCMOS camera.

The dynamic continuous measurements showcasing the binding of the csDNA, seen in Figure 4 of the manuscript, were performed under identical conditions in a petri dish. A few microlaser positions were selected using NIS-Elements software (Nikon), typically 5–10 microlasers, and their spectra were acquired at intervals of 5 or 10 seconds. A triggering mechanism and a shutter positioned in front of the diode laser were employed to time the acquisition and minimise photobleaching. After collecting at least five spectra per microlaser, csDNA was added to the buffer solution using a pipette to achieve a final concentration of at least 10 pM, corresponding to a minimum 10-fold excess. The microlasers were then monitored for approximately 20 minutes, and the resulting spectra were analyzed.

### 2.7. Simulation of microlaser emission spectra

The resonant TM and TE modal positions of the microlaser in Figure 2b-c were derived using an asymptotic expansion of the Mie solution[1]. Simulation parameters included a microlaser diameter of 11.8 µm, an internal refractive index of 1.590, and an emission range of 510–530 nm, with the external refractive index specified in the figure. Spectral peaks were modelled as Gaussian functions with an amplitude of 1 and an FWHM of 70 pm, matching the spectrometer's resolution.

## 3. DNA hybridization sensing

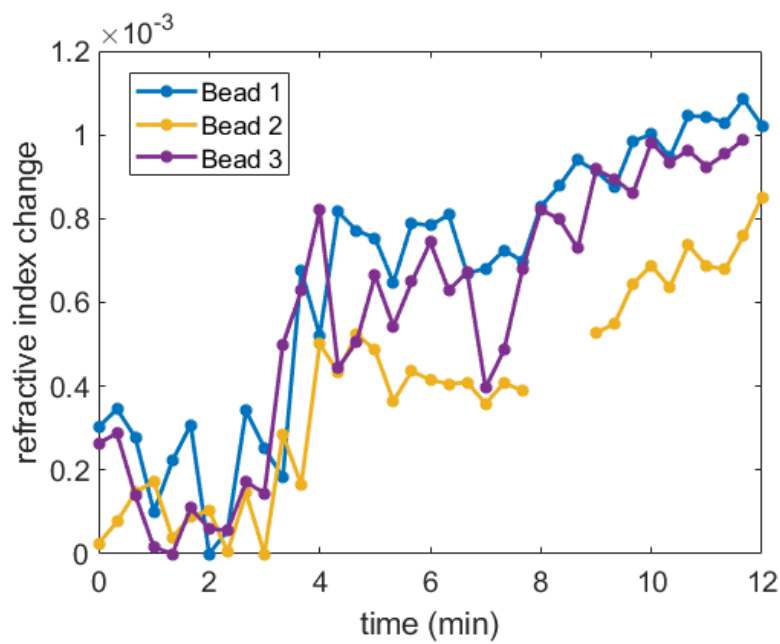

Figure S4- Transient detection of DNA hybridization on three microlasers through refractive index changes, calculated from the microlaser spectra, microlasers were measured consecutively and solution of csDNA was added at time ~3 min.

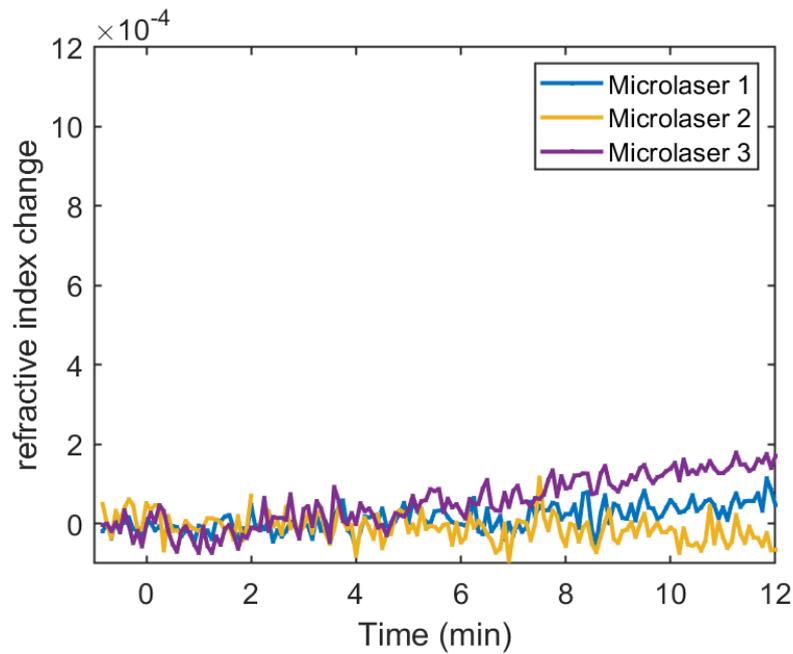

Figure S5-The refractive index at the surface of three carboxylated microlasers was monitored over time using their spectra. A solution of csDNA was added at 0 minutes. No significant changes were observed, consistent with an absence of ssDNA on the surface.

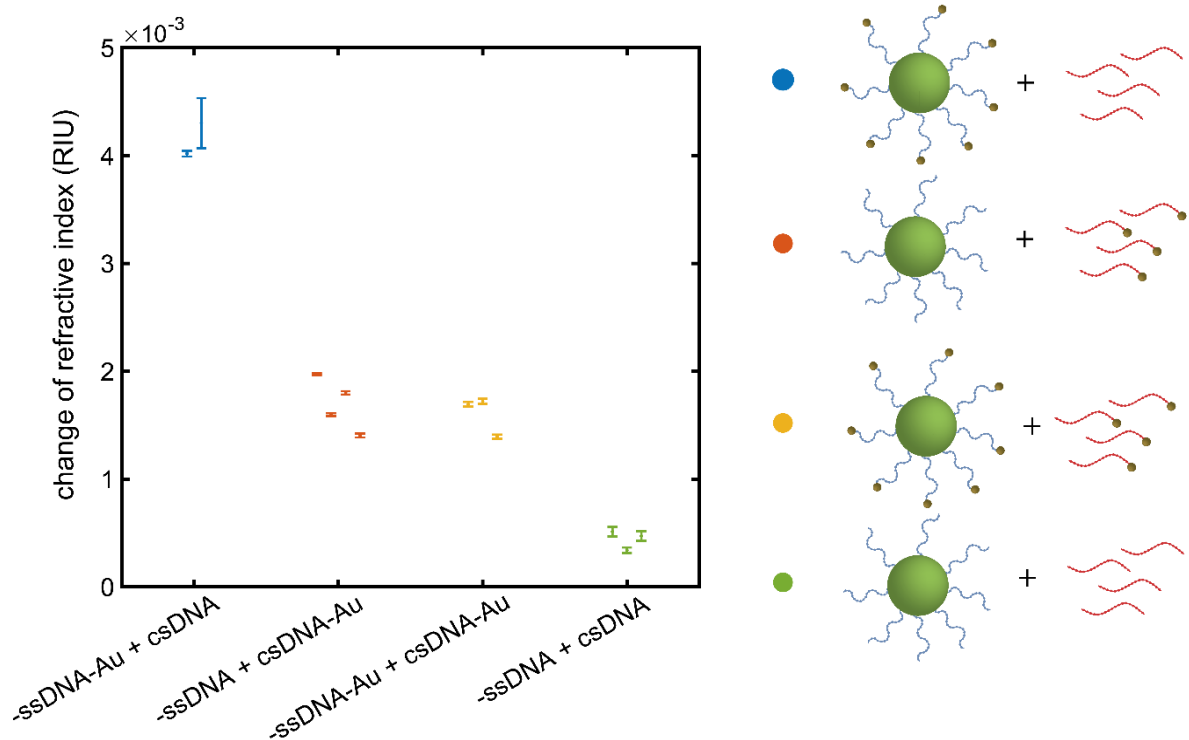

Figure S6- Changes in refractive index measured through transient detection of DNA hybridization, calculated from the microlaser spectra. The microlasers across four sample conditions. In blue, the data represents the case where the ssDNA is functionalized with gold, and csDNA (without gold) is added, as shown in Figure 4. In orange, the condition illustrates ssDNA on the microlasers without gold, while the added csDNA is functionalized with gold. In yellow, both the ssDNA on the microlasers and the added csDNA are functionalized with gold. Finally, in green, the data correspond to Figure

S3, where neither the ssDNA nor the csDNA is functionalized with gold, as suggested in the schematics on the right-hand side. Each data point represents a single micro-laser and the error bars are standard error of the difference of means error for 5-10 points before and after hybridization.

## 4. Prediction of Binding Energies:

The prediction of the lowest free energy structure of two interacting sequences was done with RNAstructure (Reuter, J. S., & Mathews, D. H. (2010). RNAstructure: software for RNA and DNA secondary structure prediction and analysis. BMC Bioinformatics. 11,129). The tool was set to DNA for nucleic acid type. The free energy of HP33 was calculated at room temperature (293 K) and compared with the energy of hybridization with csDNA.

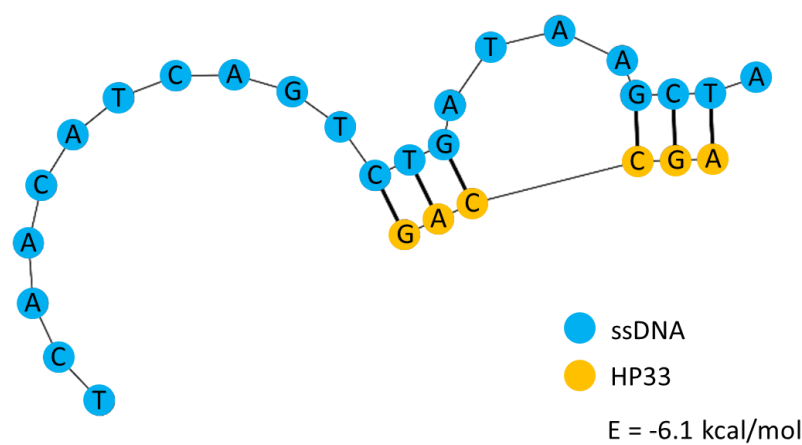

Figure S7: Predicted lowest free energy structures of hybridization of ssDNA with HP33 (top).

Table S3: Predicted energies for hybridization as shown in figures 2 and 3.

| DNA Strand 1 | DNA Strand 2 | Energy [kcal/mol] |
|---|---|---|
| ssDNA ● | csDNA ● | -32.6 |
| ssDNA | HP33 ● | -6.1 |

The calculated energy values match with experimental data, as we observed a strong bond csDNA compared to a weaker hairpin construct. HP33 has a still high free energy of -6.1 kcal/mol, which keeps it bound to the ssDNA at room temperature. A thoughtful design of the hairpin is needed, as every additional bound base pair increases the energy needed to remove it. At increased temperature this energy gets weakened. At 60°C, an opening of the hairpin construction with a decrease of the free energy to -0.6 kcal/mol was predicted. After substitution at elevated temperatures, the hairpin cannot rebind to the ssDNA, even though the measurements were done at room temperature, as the csDNA covers the ssDNA completely.

## 5. UV Melting Curve Experiments:

The final used DNA stands were tested in DNA UV melting experiments. The hybridization and cleavage of DNA can be detected using temperature controlled UV/Vis spectroscopy by measuring the temperature dependent change of absorption at 260 nm. For the experiments, UV/Vis spectroscopy was performed using a *Jasco V-730 spectrophotometer* with an *ETCS-761 Peltier thermostatic single position cell holder*. The used cuvettes were from *Starna Scientific* and had 10 mm path length and 160 µl volume. The data were analysed using *VWIS-959 temperature-Scan and DNA-melting curve analysis*.

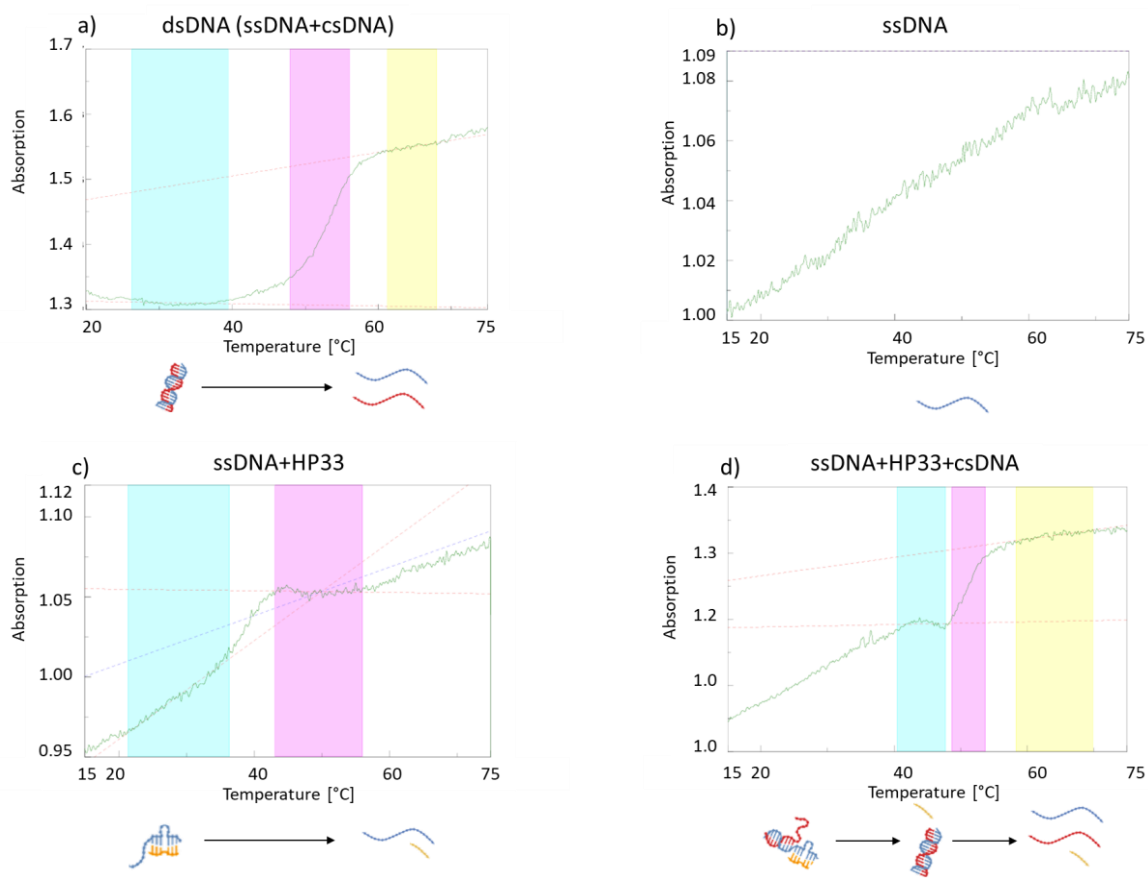

Figure S8: Measured melting curves of a) the dsDNA built by hybridization of ssDNA and csDNA, b) ssDNA, c) ssDNA with HP33 and d) the ssDNA that was first hybridized with HP33 followed by the addition of csDNA to the sample.

The UV/Vis measurements were performed in 0.01 M phosphate buffered saline (PBS), which has an equal salt concentration to the storage buffer used in the lasing experiments and mimics the sodium concentrations of intracellular medium[2, 3]. For sample preparation of dsDNA, ssDNA and ssDNA+HP33, 180 µl of the 0.01 M PBS buffer were given into an Eppendorf tube followed by 7.5 µl of a 100 µM solution of the corresponding DNA samples to yield a 4 µM final concentration. The samples were annealed at 95 °C for 5 min and slowly cooled to room temperature. For the ssDNA+HP33+csDNA, the same procedure was used, however the 7.5 µl of a 100 µM csDNA were added after the annealed ssDNA+HP33 mixture was cooled to room temperature. The samples were subsequently transferred into cuvettes. A blank sample of the PBS buffer was measured over a temperature gradient from 15 °C to 75 °C, and the data fitted by this blank. The melting curve of the dsDNA (Figure S8a) shows the typical shape of a hybridized DNA system. The absorption is linear at the beginning when all of the DNA is double stranded. Then an increase of the absorption can be seen, followed by a second plateau which indicates the absorption of the two single strands. The melting temperature ($T_m$) determines the point in this graph, where 50% of the DNA is present as ssDNA and 50% as dsDNA. This can be calculated via the three regions method ($S_1+S_2$) of the melting point analysis program and was determined to be 51.9 °C for the ssDNA+csDNA system. The second graph (Figure S8b) shows the curve of ssDNA and follows a flat slope, which fits literature results[3]. The melting curve of the HP33 system (Figure S8c) indicates a slope when the hairpin is bound to the ssDNA, followed by a plateau when both strands are cleaved and exist as single strands. The melting temperature can be determined applying the least squares method to obtain a $T_m$ of 39.4 °C. This result matches our experimental value of the hairpin substitution experiment described in the main manuscript in Figure 5c-e, which indicates a partial removal of the hairpin at 37 °C and a full substitution at 60 °C and confirms this experiment. Additionally, the hairpin substitution experiment was repeated as UV/Vis experiment, sample preparation was done as described above. The melting curve (Figure S8d) shows a slope as detected for the ssDNA+HP33 system. The $T_m$ was determined to be 37 °C by applying the method of the least squares. This slope was followed by a short plateau with a little turning point. This indicates the complete substitution of the HP33 by csDNA followed by a steep absorption increase which indicates the melting of the double strand towards single strands. This transition shows a $T_m$ of 51.8 °C which matches with the dsDNA measurement and reconfirms our results from the refractive index measurements of the hairpin experiments. The normalized experimental data are shown in Figure S9.

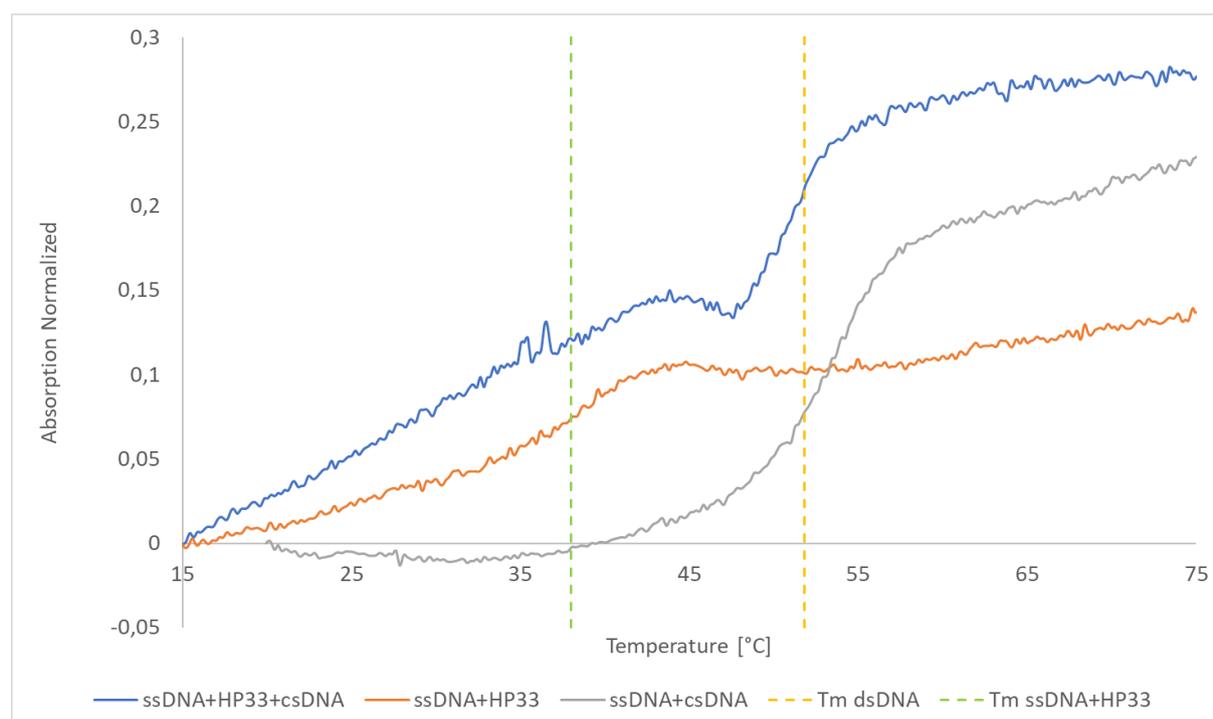

Figure S9: Melting curve diagram of the experimental melting curves from dsDNA, ssDNA+HP33 and ssDNA+HP33+dsDNA. The values are normalized. The calculated average melting temperatures were shown for the cleavage of HP33 ($T_m$ = 38 °C) and for dsDNA ($T_m$ = 51.8 °C).